\documentclass{aa}  

\usepackage{graphicx}
\usepackage{txfonts}
\usepackage{longtable}
\usepackage{rotating}
\usepackage{natbib}
\usepackage{graphics}
\usepackage{psfrag}
\usepackage{amssymb}
\usepackage{xspace}
\usepackage{color}
\usepackage{mathtools}
\bibpunct{(}{)}{;}{a}{}{,}
\def\Re{\ensuremath{R_{\oplus}}}
\def\Me{\ensuremath{M_{\oplus}}}

\def\kms{$\rm km\,s^{-1}$}

\def\planet{$\pi$\,Men\,c}
\def\star{$\pi$\,Men}
\begin{document}

\title{Three-dimensional hydrodynamic simulations of the upper atmosphere of $\pi$\,Men\,c: comparison with Ly$\alpha$ transit observations}
\subtitle{}

\titlerunning{Three-dimensional hydrodynamic simulations of the upper atmosphere of $\pi$\,Men\,c}
\authorrunning{Shaikhislamov et al.}

\author{I. F. Shaikhislamov\inst{1}	\and
	L. Fossati\inst{2}		\and
	M. L. Khodachenko\inst{2,1,3}	\and
	H. Lammer\inst{2}		\and
	A. Garc\'ia Mu\~noz\inst{4}	\and
	A. Youngblood\inst{5}		\and
	N. K. Dwivedi\inst{2}		\and
	M. S. Rumenskikh\inst{1}
}
\institute{
Institute of Laser Physics, SB RAS, Novosibirsk 630090, Russia; Institute of Astronomy, Russian Academy of Sciences, Moscow 119017, Russia\\
\email{ildars@ngs.ru}
\and 
Space Research Institute, Austrian Academy of Sciences, Schmiedlstrasse 6, A-8042 Graz, Austria
\and
Institute of Astronomy, Russian Academy of Sciences, 119992, Moscow, Russia
\and
Zentrum f\"ur Astronomie und Astrophysik, Technische Universit\"at Berlin, Hardenbergstrasse 36, D-10623, Berlin, Germany
\and
Laboratory for Atmospheric and Space Physics, 1234 Innovation Drive, Boulder, CO 80303, USA
}

\date{}

 
\abstract
{\planet\ is the first planet discovered by TESS. It orbits a bright, nearby star and has a relatively low average density, making it an excellent target for atmospheric characterisation. The existing planetary upper atmosphere models of \planet\ predict significant atmospheric escape, but Ly$\alpha$ transit observations have led to a non-detection of hydrogen escaping from the planet.}
{We aim at constraining the conditions of the wind and high-energy emission of the host star reproducing the non-detection of Ly$\alpha$ planetary absorption.}
{We model the escaping planetary atmosphere, the stellar wind, and their interaction employing a multi-fluid, three-dimensional hydrodynamic code. We assume a planetary atmosphere composed of hydrogen and helium. We run models varying the stellar high-energy emission and stellar mass-loss rate, further computing for each case the Ly$\alpha$ synthetic planetary atmospheric absorption and comparing it with the observations.}
{We find that a non-detection of Ly$\alpha$ in absorption employing the stellar high-energy emission estimated from far-ultraviolet and X-ray data requires a stellar wind with a stellar mass-loss rate about six times lower than solar. This result is a consequence of the fact that, for \planet, detectable Ly$\alpha$ absorption can be caused exclusively by energetic neutral atoms, which become more abundant with increasing the velocity and/or the density of the stellar wind. By considering, instead, that the star has a solar-like wind, the non-detection requires a stellar ionising radiation about four times higher than estimated. This is because, despite the fact that a stronger stellar high-energy emission ionises hydrogen more rapidly, it also increases the upper atmosphere heating and expansion, pushing the interaction region with the stellar wind farther away from the planet, where the planet atmospheric density that remains neutral becomes smaller and the production of energetic neutral atoms less efficient.}
{Comparing the results of our grid of models with what is expected and estimated for the stellar wind and high-energy emission, respectively, we support the idea that the atmosphere of \planet\ is likely not hydrogen-dominated. Therefore, future observations shall focus on looking for planetary atmospheric absorption at the position of lines of heavier elements, such as He, C, and O.}
\keywords{Hydrodynamics -- Planets and satellites: atmospheres --
Planets and satellites: physical evolution -- Planets and
satellites: individual: $\pi$\,Men\,c}

\maketitle
%
%
\section{Introduction}\label{sec:introduction}
The Kepler satellite revealed that planets are ubiquitous in the Milky Way \citep{coughlin2016,thompson2018}. However, most of the planets detected by Kepler lie too far away from us to enable measuring their masses and characterising their atmospheres. To obviate this problem, the TESS mission \citep{ricker2015}, and in the near future the PLATO mission \citep{rauer2014}, look for transiting planets orbiting bright, nearby stars. The first planet detected by TESS, \planet, demonstrates the success of this strategy \citep{gandolfi2018,huang2018}. \planet\ orbits a bright ($V$\,=\,5.65\,mag), nearby (d\,$\approx$\,18.3\,pc) G0\,V star previously known to host a long-period sub-stellar companion ($\pi$\,Men\,b). 

Thanks to the stellar brightness, both planetary mass and radius have been measured with high precision obtaining a mass of 4.52$\pm$0.81\,\Me\ and a radius of 2.06$\pm$0.03\,\Re\ \citep{gandolfi2018}. These measurements indicate that \planet\ is a super-Earth with a bulk density of about 2.8\,g\,cm$^{-3}$, suggesting that the planet may host a significant atmosphere, possibly water-rich, which would become hydrogen dominated in the upper layers following the dissociation of water \citep{garcia2020}. Preliminary one-dimensional (1D) hydrodynamic simulations \citep{kubyshkina2018} of a supposedly hydrogen-dominated atmosphere showed that such an atmosphere may be subject to strong escape of the order of 1.2$\times$10$^{10}$\,g\,s$^{-1}$, corresponding to roughly 1\% of the planetary mass per Gyr \citep{gandolfi2018}. Such a strong escape together with a hydrogen-dominated atmosphere would imply the presence of an extended gaseous envelope that could be directly probed by Ly$\alpha$ transit observations \citep[e.g.,][]{vidal2003}.

\citet{garcia2020} presented the results of one Ly$\alpha$ transit observation carried out with the STIS spectrograph on board HST. The data led to a clear detection of the stellar Ly$\alpha$ line, but of no planetary absorption. They obtained 1$\sigma$ upper limits for the planet-to-star radius ratio at the wavelengths covered by the Ly$\alpha$ line ($R_{\rm p,Ly\alpha}$/$R_{\rm star}$) of 0.13 and 0.12 in the [$-$215,$-$91]\,km\,s$^{-1}$ and [+57,+180]\,km\,s$^{-1}$ velocity ranges, respectively. Furthermore, the detection and reconstruction of the stellar Ly$\alpha$ line enabled them to constrain the stellar high-energy (X-ray\,+\,EUV; 5--912\,\AA; hereafter XUV) flux to 1350\,erg\,cm$^{-2}$\,s$^{-1}$ at the position of \planet\  (i.e., about 6\,erg\,cm$^{-2}$\,s$^{-1}$ at 1\,AU, which is close to the solar value). This estimate also considered the results of \citet{france2018} obtained from an HST/COS far-ultraviolet spectrum and those of \citet{king2019} obtained from archival X-ray observations. In particular, \citet{france2018} and \citet{king2019} derived stellar XUV fluxes at the distance of the planet of 1060\,erg\,cm$^{-2}$\,s$^{-1}$ and 1810\,erg\,cm$^{-2}$\,s$^{-1}$, which are within a factor of $\approx$1.3 from that given by \citet{garcia2020}.

Further to presenting the HST observations, \citet{garcia2020} showed the results of 1D hydrodynamic simulations, accounting for hydrogen and oxygen (photo)chemistry, of the planetary upper atmosphere aiming at reproducing the lack of planetary Ly$\alpha$ absorption. They showed that the mass-loss rate is somewhat sensitive to the atmospheric bulk composition, but that a water-dominated atmosphere could comply with the Ly$\alpha$ non-detection, because the presence of a large amount of oxygen would reduce the extension of the atmosphere, which would also turn from mostly neutral to mostly ionised at a low enough altitude to make neutral hydrogen not detectable at Ly$\alpha$ during transit. \citet{vidotto2020} presented 1D hydrodynamic simulations of a stellar-planetary wind interaction proposing that the non-detection of hydrogen absorption during transit may be due to the confinement below the sonic point of the planetary atmosphere by the stellar wind.
\begin{figure*}[ht!]
\centering
\includegraphics[width=7.cm]{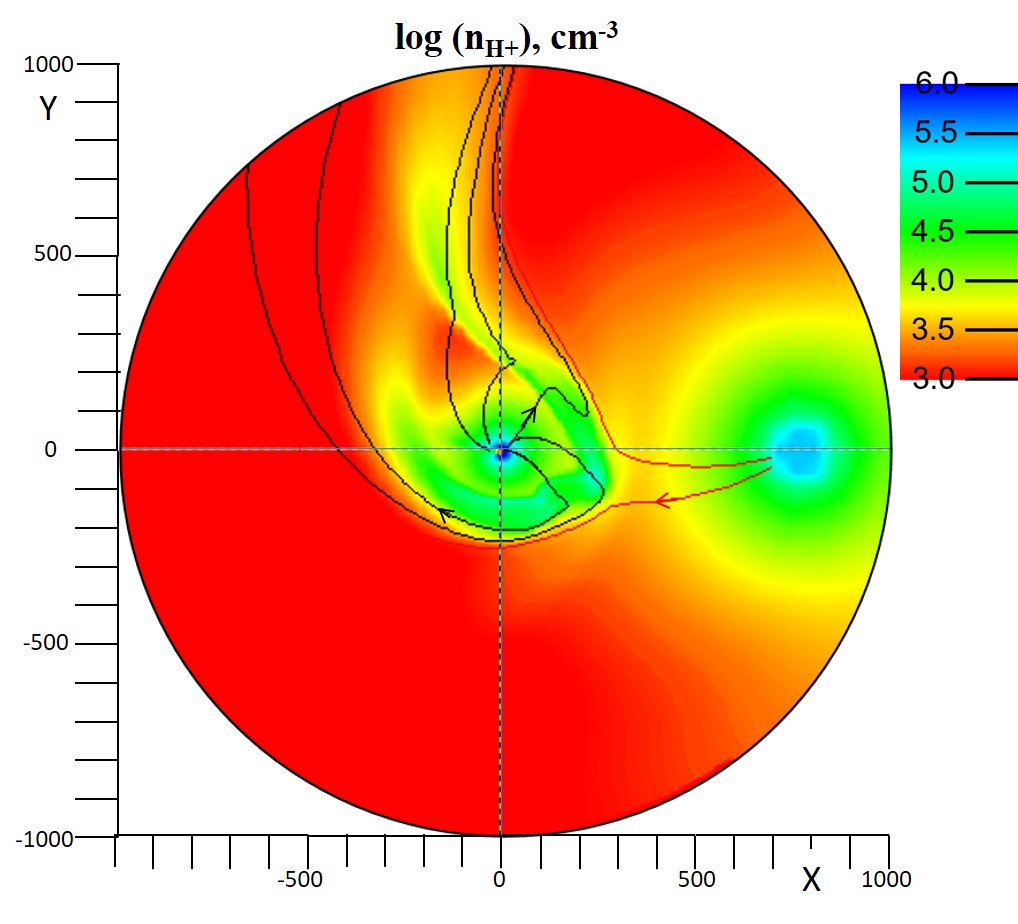}
\hspace{1.5cm}
\includegraphics[width=7.cm]{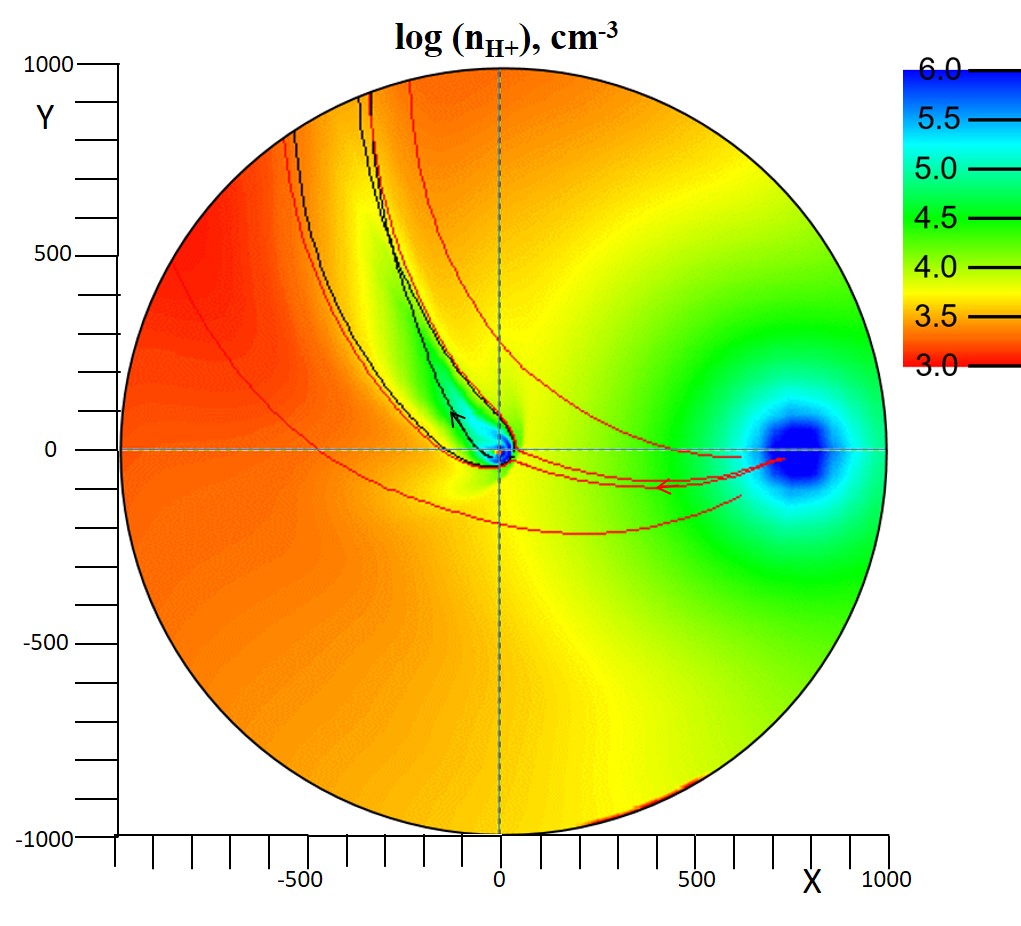}
\vspace{0.2cm}
\includegraphics[width=7.cm]{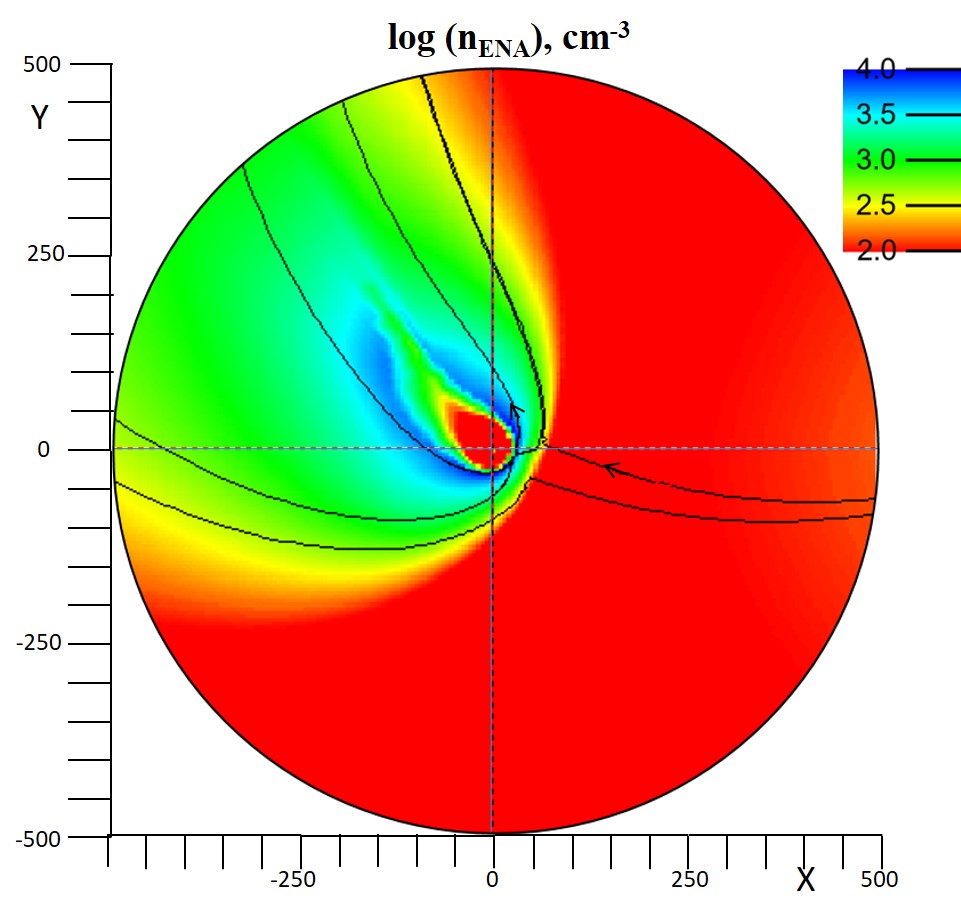}
\hspace{1.5cm}
\includegraphics[width=7.cm]{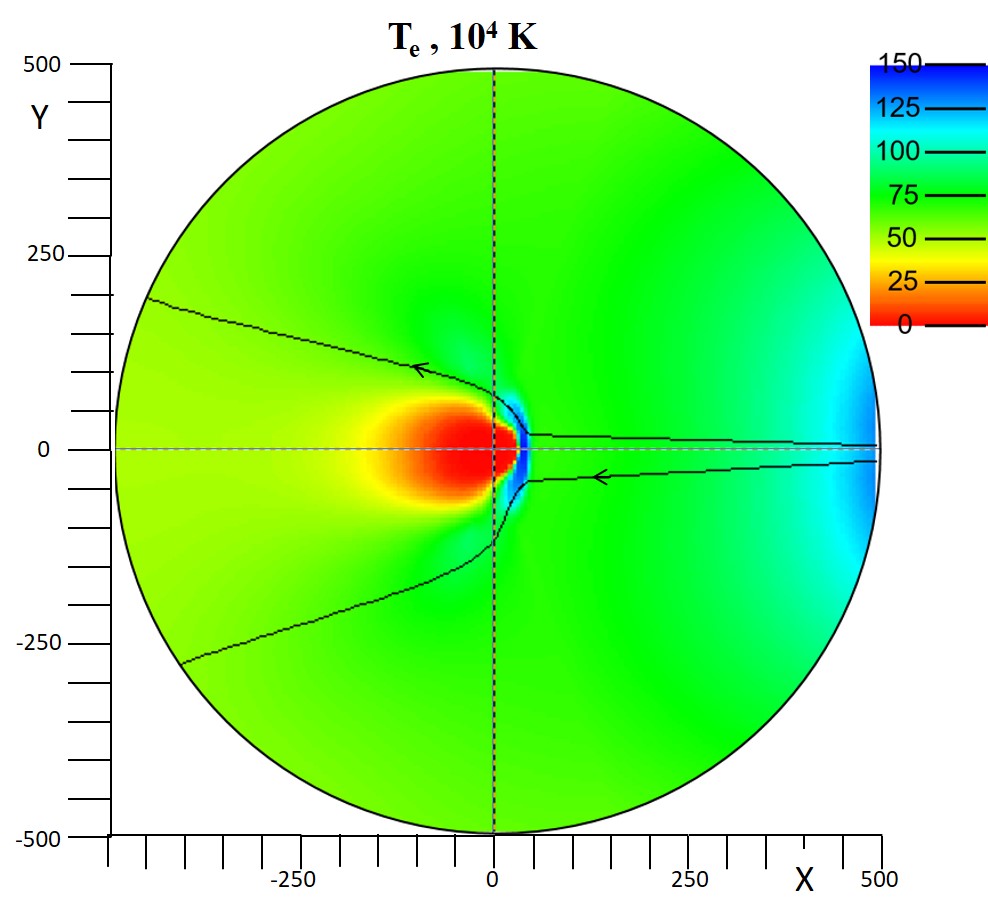}
\caption{Top left: proton density distribution in the ecliptic plane across the whole simulated domain computed with a stellar XUV flux at 1\,AU of 6\,erg\,cm$^{-2}$\,s$^{-1}$, a stellar mass-loss rate of 10$^{11}$\,g\,s$^{-1}$ (i.e., weak SW), and SW terminal velocity of 400\,km\,s$^{-1}$ (i.e., velocity of 250\,km\,s$^{-1}$ and temperature of 0.65\,MK at the planetary orbit), leading to a SW density at the planetary orbit of 3$\times$10$^2$\,cm$^{-3}$. The axes are scaled in units of planetary radii and the planet sits at the center of the coordinate reference frame, while the star lies at $X$\,=\,762. Top right: same as top-left, but for a stellar mass-loss rate of 2$\times$10$^{12}$\,g\,s$^{-1}$ (i.e., moderate SW), leading to a SW density at the planetary orbit of 6$\times$10$^3$\,cm$^{-3}$. Bottom left: same as top-right, but for the density distribution of ENAs and closing-up to the position of the planet. Bottom right: same as bottom-left, but for the electron temperature in the $X-Z$ plane. In all panels, the lines with arrows indicate streamlines of the corresponding fluids.}
\label{fig:3dmaps}
\end{figure*}

We present here full three-dimensional (3D) hydrodynamic modelling of an atmosphere of \planet, composed of hydrogen and helium, and its interaction with the stellar wind. Although the model does not take into account the range of (photo)chemistry that would be needed to simulate an atmosphere containing possibly large amounts of elements heavier than helium \citep{garcia2020}, it overcomes the limitations of the 1D assumption imposed by \citet{garcia2020} and \citet{vidotto2020}. The simulations cover two values of the stellar wind velocity and a wide range of stellar XUV emission and mass-loss rate values. With this multiparameter study, we aim at identifying the physical conditions reproducing the non-detection of the planetary Ly$\alpha$ absorption. Section~\ref{sec:model} gives a brief presentation of the modelling framework, while Sect.~\ref{sec:result} presents the results. In Sect.~\ref{sec:discussion}, we discuss the results and draw the conclusions.
\section{The theoretical framework}\label{sec:model}
We employ the 3D multi-fluid hydrodynamic model of \citet{ildar2018,ildar2020} and \citet{khodachenko2019}, which is an upgraded version of the two-dimensional one presented by \citet{khodachenko2015,khodachenko2017} and \citet{ildar2016}. We present here the most relevant features of the model. The code solves numerically the hydrodynamic continuity, momentum, and energy equations for all species in the simulated multi-component flow. In this work, we consider the atmosphere to be composed by hydrogen and helium, with a helium mixing ratio of He/H\,=\,0.1 (approximately the solar He/H abundance ratio), and account for H, H$^+$, H$_2$, H$_2^+$, H$_3^+$, He, and He$^+$. Including H$_3^+$ is important, because it influences the planetary mass loss by up to 30\% \citep[see also][]{garcia2020}. The energetic neutral atoms (ENAs) generated by charge exchange between H (of the planetary wind; hereafter PW) and H$^+$ (of the stellar wind; hereafter SW) are calculated as an independent fluid, because their velocity and temperature are significantly different from those of the neutral hydrogen atoms of planetary origin. The photochemical reactions of the H-He plasma are described in \citet{khodachenko2015} and \citet{ildar2016}.

Photoionisation of both hydrogen and helium results in strong heating by the produced photoelectrons, which is the driver of the hydrodynamic outflow of the planetary atmosphere. The model derives the corresponding heating term by integrating the stellar XUV spectrum of \citet{garcia2020}. As shown, for example, by \citet{khodachenko2019}, the heating term is computed as
\begin{equation}
\begin{multlined}
\label{eq:heating}
W_{\rm XUV} = \frac{1}{N_{\rm tot}}(\gamma_{\rm a}-1)\,\,\times\,\,n_{\rm a} [\langle(\hslash\nu - E_{\rm ion})\sigma_{\rm XUV}F_{\rm XUV}\rangle\,\, \\
-\,\,n_{\rm e} \nu_{\rm Te} (E_{21}\sigma_{21} + E_{\rm ion}\sigma_{\rm ion})]\,,
\end{multlined}
\end{equation}
where $N_{\rm tot}$ is the total density of all particles, including electrons, $\gamma_{\rm a}$ is the adiabatic specific heat ratio that we take being 5/3, $n_{\rm a}$ is the density of hydrogen atoms, $h\nu$ is the photon energy, $E_{\rm ion}$ is the hydrogen ionisation energy of 13.6\,eV, $\sigma_{\rm XUV}$ is the wavelength dependent ionisation cross section to XUV radiation, $F_{\rm XUV}$ is the XUV stellar flux at the planetary orbital distance, $n_{\rm e}$ is the electron density, $\nu_{\rm Te}$ is the terminal velocity of electrons, $E_{21}$ is the $n$\,=\,1 level to $n$\,=\,2 level hydrogen excitation energy, and $\sigma_{21}$ and $\sigma_{\rm ion}$ are the hydrogen excitation and ionisation cross sections by electron impact, respectively. Under these conditions, the photoionisation time of unshielded hydrogen atoms at the planetary orbit is about 4 hours.

The model equations are solved in a non-inertial spherical reference frame fixed at the planetary center and rotating at the same rate as the planet orbits around the star, so that the planet faces the star always with the same side. The $X$ axis connects the star and the planet, while the $Y$ axis is perpendicular to the $X$ axis and lies in the ecliptic plane. The polar-axis $Z$ is directed perpendicular to the ecliptic plane and completes the so-called tidally locked spherical reference frame. In this frame, we properly account for the non-inertial terms, namely the generalised gravity potential and Coriolis force. Since the gas beyond the planetary exobase is mostly ionised, the species in the simulation domain are collisional through the action of the Coulomb force, justifying the hydrodynamic approach \citep[e.g.,][]{debrecht2020,vidotto2020}. Furthermore, the model also takes into account radiation pressure, which we find being mostly negligible compared to other forces \citep[e.g.,][]{murray2009,khodachenko2017,khodachenko2019,debrecht2020}. We tested this by computing models with twice and ten times larger stellar Ly$\alpha$ emission compared to the baseline of 5.6\,erg\,s$^{-1}$\,cm$^{-2}$ at 1\,AU. We obtained that only the model with the highest Ly$\alpha$ flux affects the Ly$\alpha$ planetary absorption profile, but only at velocities between -100 and 0\,km\,s$^{-1}$, which are anyway too low to be detectable by the observations and contaminated by interstellar medium (ISM) absorption and geocoronal emission \citep{garcia2020}.

The fluid velocity at the planetary surface is taken to be zero. Furthermore, to keep the number of grid points in the model small enough to be manageable by the numerical code, the radial mesh is highly non-uniform, with the grid step increasing linearly from the planetary surface. This allows us to resolve the highly stratified upper atmosphere of the planet, where the required grid step is as small as $\Delta$$r$\,=$R_{\rm p}$/400, where $r$ is the radial distance from the center of the planet and $R_{\rm p}$ the planetary radius. As initial state, we take a fully neutral atmosphere in barometric equilibrium composed by H$_2$ and He. 

For all simulations, we considered the system parameters of \citet{gandolfi2018} that we reproduce in Table~\ref{tab:system_parameters}. At the inner boundary of the simulation domain, at $r$\,=\,$R_{\rm p}$, we set a temperature and a pressure of 1000\,K and 0.05\,bar, respectively. The former is close to the planetary equilibrium temperature obtained considering zero albedo (see Table~\ref{tab:system_parameters}), while the latter is the pressure at which the planetary atmosphere is optically thick to photons with a wavelength longer than of 10\,\AA\  (i.e., the blue edge of the XUV range).
\begin{table}[h!]
\caption{Adopted system parameters of \star\ and \planet\ from \citet{gandolfi2018}.}
\label{tab:system_parameters}
\begin{center}
\begin{tabular}{l|c}
\hline
\hline
Parameter & Value \\
\hline
Stellar mass, $M_{\rm s}$ [$M_{\odot}$]   & 1.02   \\
Stellar radius, $R_{\rm s}$ [$R_{\odot}$]   & 1.10   \\
Semi-major axis of planet c, $a$ [AU] & 0.06702 \\
Planetary mass, $M_{\rm p}$ [$M_{\rm Earth}$] & 4.52    \\
Planetary radius, $R_{\rm p}$ [$R_{\rm Earth}$] & 2.06   \\
Planetary equilibrium temperature, T$_{\rm eq}$, [K] & 1147 \\
\hline
\end{tabular}
\end{center}
\end{table}

The model also incorporates self-consistently the flow of the SW plasma, which we consider being composed by protons, in the way described by \citet{khodachenko2019} and \citet{ildar2020}, enabling one to model the whole system. Thus, the simulation domain has one further boundary at the base of the stellar corona. At distances from the stellar center shorter than 20 stellar radii, the SW is accelerated by an empirical heating term, derived from an analytical 1D polytropic Parker-like model \citep{keppens1999}, that we compute as 
\begin{equation}
\label{eq:heating_star}
W_{\rm SW} = (\gamma_{\rm a} - \gamma_{\rm p})\,\,\times\,\,T_{\rm p}(r_{\rm s})\,\,\times\,\,div\,V_{\rm p}(r_{\rm s})\,. 
\end{equation}
In Eq.~\ref{eq:heating_star}, $\gamma_{\rm a}$ is the adiabatic specific heat ratio that we take being 5/3 to correctly model shocks and $r_{\rm s}$ is the distance from the center of the star, while $\gamma_{\rm p}$, $T_{\rm p}$, and $V_{\rm p}$ are respectively the polytropic index, temperature, and velocity obtained from the polytropic solution. The simulated stellar wind is isotropic in space, stationary in time, and in good agreement with the Parker analytical solution.

We further employ the simulations to compute the absorption at the position of the Ly$\alpha$ line following \citet{ildar2018}. This procedure has already been successfully employed to generate synthetic observations for the hot Jupiter HD\,209458\,b \citep{khodachenko2017,ildar2018,ildar2020} and the warm Neptune GJ\,346\,b \citep{khodachenko2019}, providing physically reasonable and self-consistent interpretation of Ly$\alpha$ transit observations, as well as of transit observations of resonance lines of minor species (e.g., C, O, Si).
\section{Results}\label{sec:result}
%
The velocity of the PW driven by XUV heating reaches moderate values of about 10\,km\,s$^{-1}$, which are by far not enough to produce Doppler shifted absorption beyond the wavelengths strongly contaminated by ISM absorption and geocoronal emission. Therefore, absorption in the wings of the Ly$\alpha$ line during transit can be caused only by ENAs generated during the interaction of the PW with the SW.

$\pi$\,Men is an $\approx$5\,Gyr old solar-type star \citep{gandolfi2018}, which is therefore likely to have solar-like wind properties. To explore how strong the Ly$\alpha$ absorption might be under different physical conditions, we simulate the PW-SW interaction considering two winds typical of solar-like plasma, namely a fast and a slow wind of terminal velocities $V_{\rm sw,\infty}$\,=\,800\,km\,s$^{-1}$ and 400\,km\,s$^{-1}$, respectively, further varying the stellar mass-loss rate, hence SW density, by about an order of magnitude. Another parameter directly affecting the Ly$\alpha$ absorption is the stellar XUV emission, which we also vary by about an order of magnitude.

Figure~\ref{fig:3dmaps} presents simulation results for two cases. The first case is for a very weak SW with a density ten times smaller than solar. The stellar mass-loss rate in this case is of 10$^{11}$\,g\,s$^{-1}$. Under these conditions, the SW, which is still subsonic at the planetary orbital separation, diverts the planetary outflow only far from the planet, at an $X$\,=\,$r$/$R_{\rm p}$ value of $\approx$200--300. Therefore, the SW-PW interaction occurs in a low-density region of the planetary atmosphere, leading to a small ENA production. The second case is for a moderate SW, namely with a stellar mass-loss rate of 2$\times$10$^{12}$\,g\,s$^{-1}$ \citep[the average solar mass-loss rate is about 2.5$\times$10$^{12}$\,g\,s$^{-1}$; e.g.,][]{phillips1995}. This second case is significantly different from the previous one, because the SW pressure is enough to stop the planetary outflow close to the planet, at $X$ $\approx$20--30, redirecting the planetary escaping material towards the tail, further generating a large amount of ENAs. Because of its supersonic nature, this interaction generates a bow shock, as evinced by the temperature distribution (Fig.~\ref{fig:3dmaps}). The ENAs, which produce significantly high velocity absorption in the blue wing of the Ly$\alpha$ line, are generated inside the bow shock region.

\begin{figure}
\centering
\vspace{-0.3cm}
\includegraphics[width=\hsize,clip]{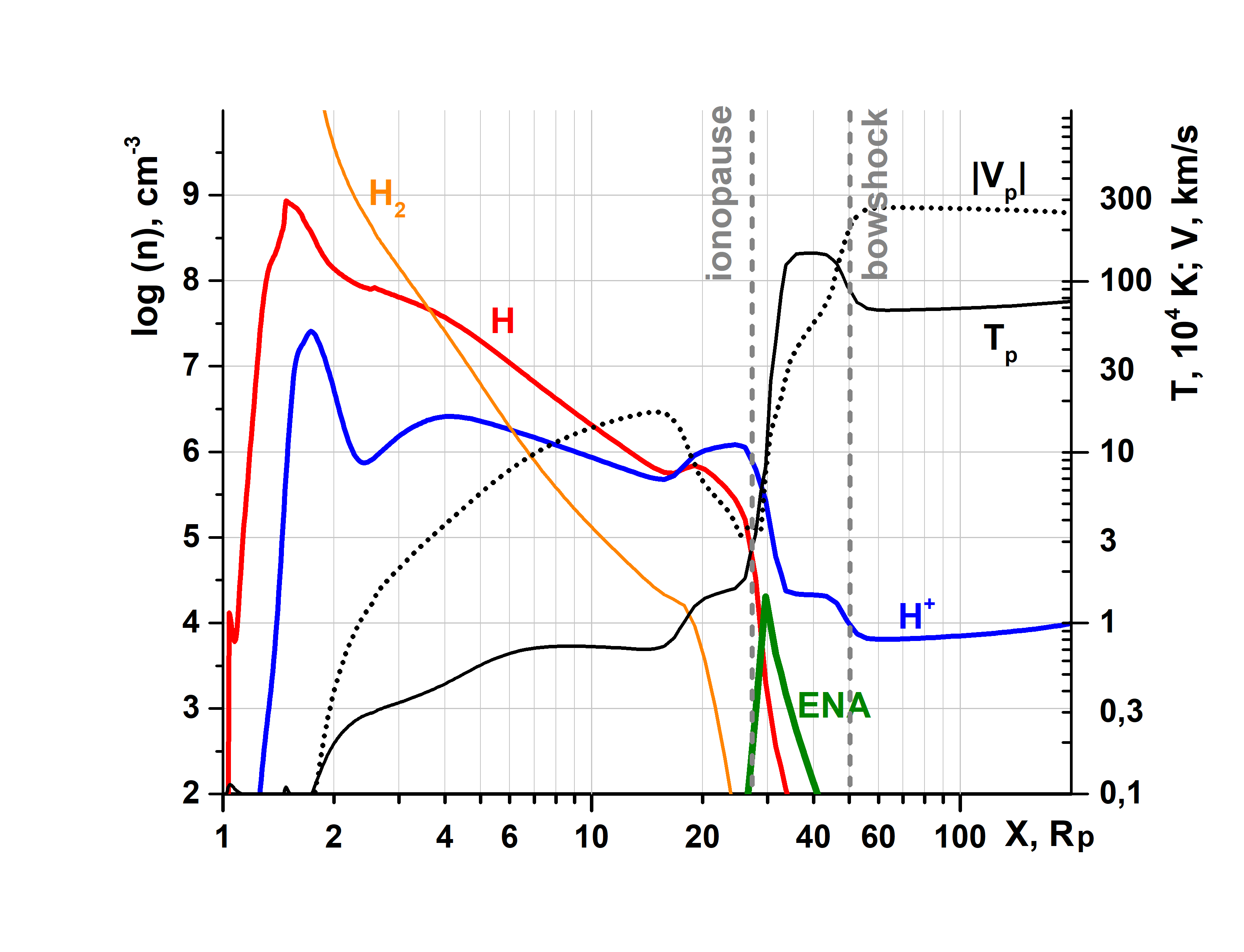}\vspace{-0.5cm}
\caption{Distribution profiles of major species along the planet-star line obtained in the simulation with the moderate SW (i.e., top-right and bottom panels of Fig.~\ref{fig:3dmaps}). The left axis is for the density (in log[cm$^{-3}$]) of H$_2$ (orange line), H (red line), H$^+$ (blue line), and ENAs (green line). The right axis is for the proton temperature (in 10$^4$\,K; black solid line) and velocity (in km\,s$^{-1}$; dotted line). The vertical dashed lines indicate the approximate positions of the ionopause and of the bow shock. The planet lies at $X$\,=\,0, while the star is located to the right.}
\label{fig:details}
\end{figure}

Figure~\ref{fig:details} presents the details of the simulated distribution of species computed for the moderate SW (i.e., top-right and bottom panels of Fig.~\ref{fig:3dmaps}) along the star-planet line. It shows that the simulation domain is split by the shock into two regions, one dominated by the escaping planetary material and one dominated by SW material. The supersonic SW remains undisturbed until the bow shock region, where it experiences a sharp deceleration, compression, and heating. The region dominated by planetary material, instead, is characterised by a relatively low temperature and velocity, though before the ionopause the PW is supersonic. At the ionopause, the normal component of the proton velocity goes to zero. However, the neutral PW particles penetrate into the shock region, where they are rapidly ionised by the hot electrons and XUV radiation. In the shocked region, adjacent to the ionopause, the energetic SW protons charge exchange with planetary atoms producing ENAs. For this simulation, we obtained a planetary mass-loss rate of about 2$\times$10$^{10}$\,g\,s$^{-1}$, which is in good agreement with previous estimates based on 1D simulations \citep{gandolfi2018,garcia2020}; we obtain roughly the same planetary atmospheric mass-loss rate (within a few \%) for all conducted simulations. We remark that, besides the difference in the geometry of the simulations, the mass-loss rates of \citet{gandolfi2018} were obtained accounting only for H and those of \citet{garcia2020} for H and O, while our simulations consider H and He.

The result that models with different assumptions and geometries lead to similar planetary mass-loss rates is found also for other planets \citep[see, for example, Table 2 of][]{kubyshkina2018b}. This may be because the physical properties (mainly density and velocity) of the gas at the sonic point are robust against the typical assumptions taken in the different codes, though detailed comparisons of the model outputs would be necessary to confirm this. However, the most interesting result is possibly the similarity of mass-loss rates computed by 1D and 3D models, where the former are integrated across the whole planet (i.e., 1D mass-loss rates multiplied by 4$\pi$). Indeed, although one might expect that mass loss would be strongly reduced on the night side, the large size of the upper atmosphere of close-in planets and the redistribution of heat across it lead to similar conditions throughout the majority of the upper atmosphere.

Figure~\ref{fig:details} enables one to gather a rough estimate of the absorption at the position of the Ly$\alpha$ line. The integral of the density of ENAs along the $X$ axis (i.e., the ENA column density; hereafter NL) is equal to NL\,=\,4.5$\times$10$^4$\,$R_{\rm p}$\,=\,0.6$\times$10$^{14}$\,cm$^{-2}$, while the resonant cross-section absorption of the Ly$\alpha$ line is $\sigma$\,=\,6$\times$10$^{-14}$\,cm$^2$ \citep{khodachenko2017}. Assuming that ENAs form a shell of radius $R_{\rm abs,ENAs}$\,$\sim$\,30\,R$_{\rm p}$\,$\sim$\,0.5\,$R_{\rm star}$ around the planet (see Fig.~\ref{fig:3dmaps}), the Ly$\alpha$ absorption caused by ENAs is 0.5$^2$\,NL\,$\sigma$\,$\approx$\,0.9, meaning that the size of the planet at Ly$\alpha$ wavelengths $R_{\rm p,Ly\alpha}$/$R_{\rm star}$ is $\approx$\,1.0. 

\begin{figure}
\centering
\vspace{-0.3cm}
\includegraphics[width=\hsize]{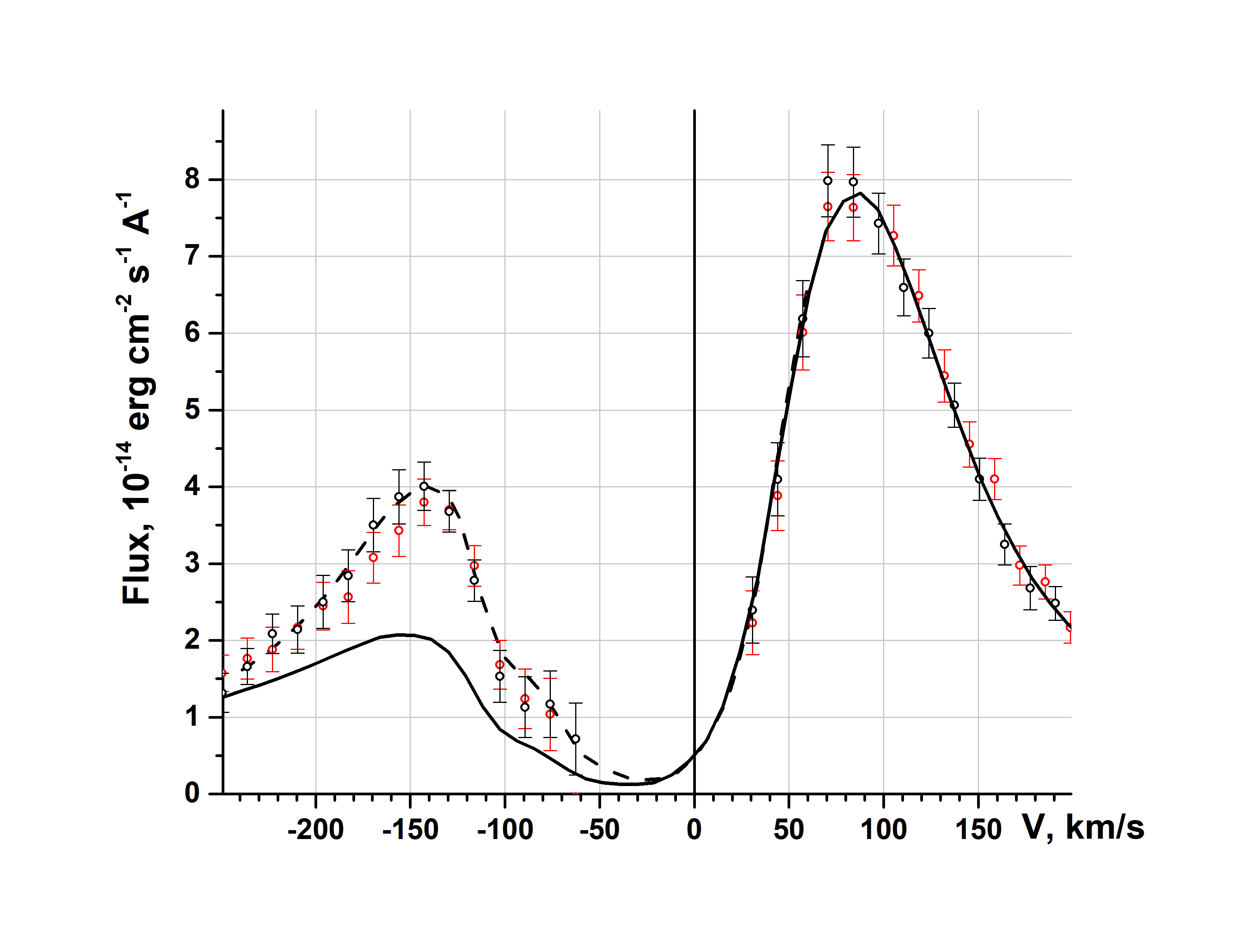}\vspace{-0.5cm}
\includegraphics[width=6.cm]{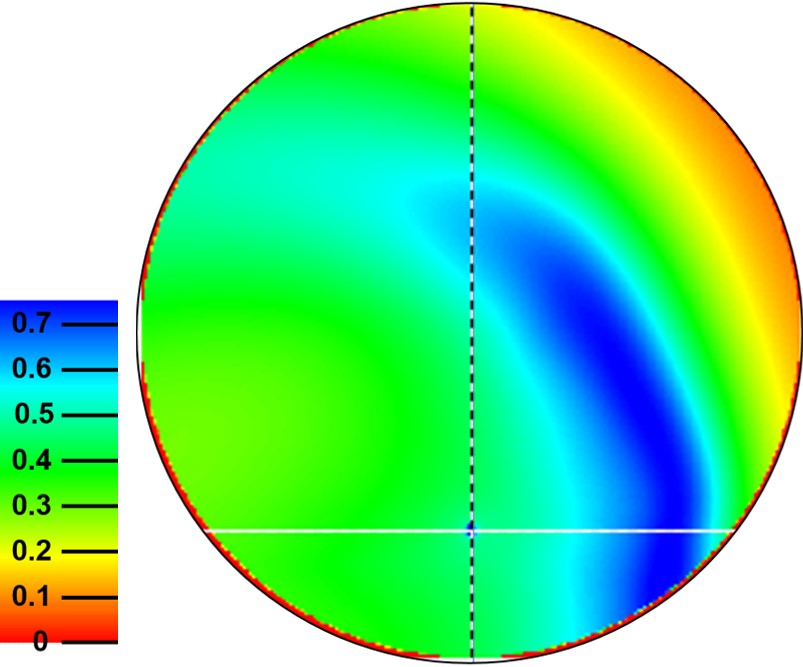}
\caption{Top: Out-of-transit (black circles) and in-transit (red circles) observed Ly$\alpha$ profiles \citep[from][]{garcia2020}. The dashed line shows the synthetic in-transit Ly$\alpha$ profile obtained from the simulation computed under the weak SW conditions (i.e., top-left panel in Fig.~\ref{fig:3dmaps}), while the solid line is for the simulation computed with the moderate SW (i.e., top-right and bottom panels of Fig.~\ref{fig:3dmaps}). Of the two simulations represented in this figure, only the weak SW case is consistent with the HST non-detection. Bottom: distribution of the Ly$\alpha$ absorption along the line of sight averaged over the blue wing of the line, in the [$-$215,$-$91]\,km\,s$^{-1}$ velocity range, as seen by a remote observer at mid-transit and considering the moderate SW. The absorption ranges between 0 and 1, where 0 means no absorption, while 1 means full absorption. The black circle at the outer boundary indicates the star and the white horizontal line shows the planetary orbital path accounting for the impact parameter (the planet moves form left to right). At mid-transit, the planet is located at the intersection of the horizontal white solid line and of the vertical dashed black line.}
\label{fig:absorption}
\end{figure}

Figure~\ref{fig:absorption} shows the actual absorption profile obtained from the simulation computed considering the weak and moderate SW (i.e., top panels of Fig.~\ref{fig:3dmaps}), and the distribution of the wavelength-integrated absorption depths (i.e., 1$-\exp^{-{\rm NL}\,\sigma}$) across the stellar disk obtained from the two simulations. The absorption profiles have been computed accounting for all neutral hydrogen particles, further considering their own velocities and temperatures. In the weak SW case, the absorption takes place mostly in the [-30,50]\,km\,s$^{-1}$ velocity range, where any planetary absorption signature is unobservable, because of contamination by ISM absorption and geocoronal airglow emission. In contrast, in the moderate SW case, the absorption is largest at higher negative velocities and comes mostly from the shocked region, as shown by the bottom panel of Fig.~\ref{fig:absorption}, which presents the distribution of the high-velocity absorption, hence from ENAs, across the stellar disk.

\begin{figure}
\centering
\includegraphics[width=9cm]{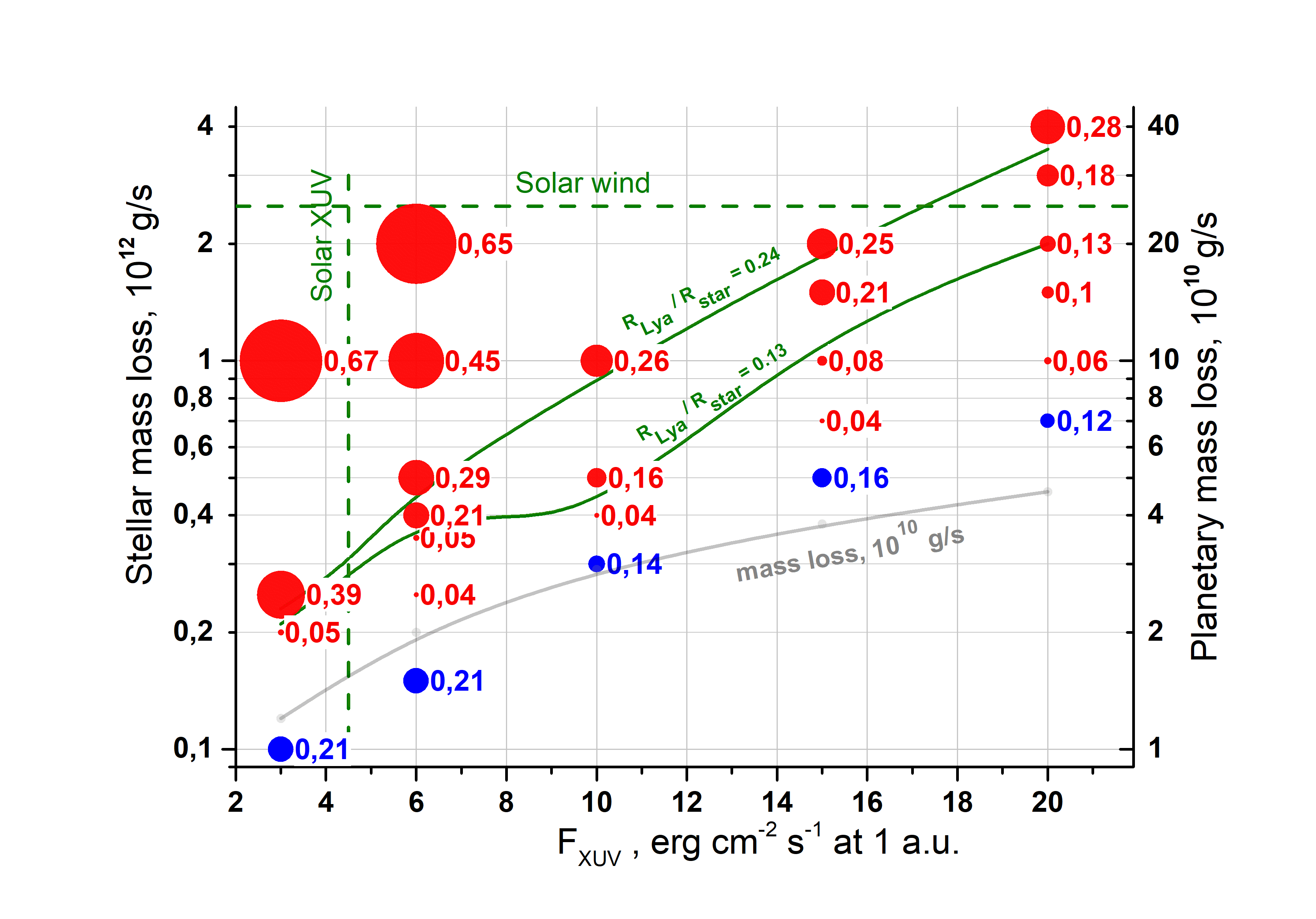}
\caption{Planetary Ly$\alpha$ absorption in $R_{\rm p,Ly\alpha}$/$R_{\rm star}$ integrated in the [$-$215,$-$91]\,km\,s$^{-1}$ velocity range as a function of stellar XUV flux (at 1\,AU; x-axis) and input stellar mass-loss rate (left y-axis). Red circles indicate the results obtained with a SW temperature and terminal velocity of 0.64\,MK and 400\,km\,s$^{-1}$, respectively. Blue circles are for a SW temperature and terminal velocity of 1.2\,MK and 800\,km\,s$^{-1}$, respectively. The size of the red circles is proportional to the planetary absorption, whose value is given beside each circle. For comparison with observations, the green solid lines show the approximate position of the 0.13 and 0.24 absorption levels with respect to the results given by the red circles. The gray curve gives the planetary mass-loss rate (right y-axis) obtained from the simulations as a function of stellar XUV flux in the case of the slow SW. The planetary mass-loss rate is very weakly dependent of the stellar mass-loss rate (see Sect.~\ref{sec:discussion}).}
\label{fig:summary}
\end{figure}

Finally, we performed a systematic study of the Ly$\alpha$ planetary absorption depth as a function of the stellar XUV flux, ranging between 3 and 20\,erg\,s$^{-1}$\,cm$^{-2}$ at 1\,AU, and of the stellar mass-loss rate, ranging between 2$\times$10$^{11}$ and 4$\times$10$^{12}$\,g\,s$^{-1}$. For this grid of models, the SW terminal velocity is kept constant at the slow SW, namely 400\,km\,s$^{-1}$. We run an additional, smaller grid considering the fast SW, namely 800\,km\,s$^{-1}$, and stellar mass-loss rates, ranging between 10$^{11}$ and 7$\times$10$^{11}$\,g\,s$^{-1}$. We remind the reader that the estimated XUV flux of $\pi$\,Men at 1\,AU is 6\,erg\,s$^{-1}$\,cm$^{-2}$, while the average solar mass-loss rate is 2.5$\times$10$^{12}$\,g\,s$^{-1}$, hence comparable to the strongest SW we considered (i.e., that with the highest stellar mass-loss rate). Figure~\ref{fig:summary} summarises the results of the systematic analysis. It presents the absorption in terms of $R_{\rm p,Ly\alpha}$/$R_{\rm star}$ integrated over the blue wing of the Ly$\alpha$ line in the [$-$215,$-$91]\,km\,s$^{-1}$ velocity range, for which \citet{garcia2020} obtained 1$\sigma$ and 3$\sigma$ upper limits of $R_{\rm p,Ly\alpha}$/$R_{\rm star}$\,=\,0.13 and 0.24, respectively. We do not consider in the analysis the red wing of the Ly$\alpha$ line, because at those velocities none of the computed models led to absorption redwards of the region contaminated by ISM absorption and geocoronal emission, in agreement with the observations.

We first analyse the results obtained from the larger grid computed considering the slow SW. Figure~\ref{fig:summary} indicates that all simulations in the upper-left quadrant (i.e., lower XUV flux and larger stellar mass-loss rate) show strong Ly$\alpha$ absorption, contrary to the observations. For each considered XUV flux value, there is a narrow band of stellar wind densities, following roughly the lower green full line in Fig.~\ref{fig:summary}, across which the planetary absorption decreases significantly. This is because a stronger SW (i.e., larger stellar mass-loss rate) will stop the expanding PW closer to the planet (see e.g. Fig.~\ref{fig:3dmaps}), where the ENA production is efficient. In contrast, a weaker SW (i.e., smaller stellar mass-loss rate) will result in inefficient ENA production and therefore the planetary absorption in the blue wing of Ly$\alpha$ comes only from natural line broadening, which amounts to $R_{\rm p,Ly\alpha}$/$R_{\rm star}$\,$\approx$\,0.04. Despite this small value, the planetary plasmasphere is as large as $\sim$10\,R$_{\rm p}$, because the SW is not strong enough to confine the planetary atmosphere.

Figure~\ref{fig:summary} shows that, in comparison to the slow SW case (i.e., SW terminal velocity of 400\,\kms), the fast SW (i.e., SW terminal velocity of 800\,\kms), even with the same stellar mass-loss rate, produces significantly higher Ly$\alpha$ absorption. This is because the pressure of the fast SW is higher, confining the planetary atmosphere inside a bow shock closer to the planet. This implies that the fast SW interacts with a denser PW, leading to a larger ENA production that generates stronger absorption in the blue wing of the Ly$\alpha$ line (see bottom panel of Fig.~\ref{fig:absorption}).
\section{Discussion and conclusion}\label{sec:discussion}
We ran 3D hydrodynamic simulations of the interaction between the expanding upper atmosphere of \planet\ and of the wind of its host star, considering a planetary atmosphere composed of H and He. We ran simulations assuming two distinct values of the SW terminal velocity (400 and 800\,km\,s$^{-1}$) and a range of stellar XUV fluxes and mass-loss rates. We find that, assuming a slow SW and for stellar XUV fluxes close to those estimated for $\pi$\,Men, the non-detection of Ly$\alpha$ absorption during transit can be reproduced by considering stellar winds significantly weaker than the average solar wind, with a density more than 6 times smaller. Reproducing the Ly$\alpha$ non-detection employing a faster SW would require an even lower SW density. We find that with a solar-like SW, fitting the Ly$\alpha$ non-detection would require an improbably higher stellar XUV flux, namely about 4 times of that estimated for $\pi$\,Men, while the highest XUV estimate is of just $\approx$1.3 times larger \citep{king2019} and uncertainties on the reconstructed XUV fluxes based on Ly$\alpha$ measurements are typically of the order of 30\% \citep{linsky2014}. This is because a higher XUV flux more rapidly ionises hydrogen, increasing the upper atmospheric heating and expansion, pushing the interaction region with the SW farther away from the planet. Furthermore, at such high stellar XUV fluxes the planet would have lost almost half of its current mass within the estimated age of the system, without accounting for the fact that the star was more active in the past. On the basis of the estimated stellar XUV flux, accounting for its uncertainty, and of evolutionary considerations, we conclude that a high stellar XUV emission is unlikely to be the cause of the non-detection of planetary Ly$\alpha$ absorption.

This result is driven by the fact that Ly$\alpha$ planetary atmospheric absorption at the velocity probed by the observations can be caused just by ENAs, which become more abundant with increasing the velocity and/or the density of the SW. Therefore, similarly to \citet{vidotto2020}, we find that a stronger (i.e., faster and/or denser) SW compresses the planetary atmosphere on the side facing the star, reducing its size. However, a stronger SW penetrates deeper into the neutral part of the expanding planetary atmosphere, increasing the density of ENAs, thus the Ly$\alpha$ absorption at the velocities probed by the observations. 

However, the set of simulations described in Sect.~\ref{sec:result} do not reproduce the case assumed by \citet{vidotto2020} in which the SW is so strong that it compresses the planetary atmosphere below the sonic point. Therefore, we run a further simulation considering the slow SW (i.e., 400\,km\,s$^{-1}$), a stellar XUV flux at 1\,AU of 6\,erg\,cm$^{-2}$\,s$^{-1}$, and a stellar mass-loss rate of 1.7$\times$10$^{13}$\,g\,s$^{-1}$ and density of 5$\times$10$^4$\,cm$^{-3}$, closely corresponding to Model A in \citet{vidotto2020}. In agreement with \citet{vidotto2020}, we find that the bow shock and ionopause lie much closer to the planet, namely at 16.5 and 10.5\,$R_{\rm p}$, respectively. We also find that the planetary flow on the day side reaches its maximum speed of 4.9\,km\,s$^{-1}$ at a distance of 6\,$R_{\rm p}$, while the sound speed is 14.4\,km\,s$^{-1}$, implying a subsonic flow.

However, we also find that the planetary mass loss rate is just 2.5\% smaller than what we obtained from the other simulations, therefore confirming that a subsonic PW does not necessarily entail a smaller mass-loss rate \citep{garcia2007}. In one-dimensional models, this is the result of partial compensation in density and velocity changes when the prescribed downstream pressure is increased. Indeed, higher pressures slow down the flow, which responds by increasing the temperature and in turn the density (through a larger scale height). Because the mass loss rate scales with the product of density and velocity, their individual changes tend to cancel out. There is a limit to this, as for instance expected if the planetary wind temperatures become very high and the plasma loses a significant amount of energy through radiation. This picture is qualitatively consistent with the parametric study conducted by \citet{christie2016}, who report a decrease in the mass-loss rate when the planetary wind becomes weaker and easier to confine by the stellar wind (see their Fig. 7). The quantitative differences with \citet{christie2016} probably arise from the different treatment of the equation of state of the gas. It is unclear if their prescription of a polytropic equation of state, which results in roughly isothermal temperature of the planetary wind, can capture the compensation effects described above.

Furthermore, we find that absorption in the blue wing of Ly$\alpha$ is still very large, namely $R_{\rm p,Ly\alpha}$/$R_{\rm star}$\,$\approx$\,0.64, which is in disagreement with the observations. \citet{vidotto2020} could not have reached this result, because of the 1D assumption and, more importantly, the lack of ENAs. Therefore, although a very strong SW is indeed able to slightly reduce the planetary mass-loss rate, it would also lead to a stronger, rather than weaker, absorption signature in the blue wing of the Ly$\alpha$ line profile.

Since estimates indicate that the wind strength of $\pi$\,Men is close to solar \citep[or even stronger;][]{vidotto2020}, our simulations clearly suggest that it is very unlikely \planet\ hosts an atmosphere dominated by hydrogen and helium, in agreement with the considerations of \citet{gandolfi2018} and \citet{garcia2020}. Therefore, we argue that, despite the rather low bulk planetary density, \planet's atmosphere cannot be strongly hydrogen-dominated and that should thus contain a non-negligible amount of heavier elements, as suggested by \citet{garcia2020}. Future observations should focus on looking for elements, such as He, C, and O, to shed more light on the planetary atmospheric composition. Furthermore, additional observations are needed to check that the available measurements have not been made at non-typical and rare stellar conditions of either very low SW density or high XUV flux.
%
\begin{acknowledgements}
I.S., M.Kh., and M.R. received support by the RSF project 18-12-00080 in the frame of which the numerical modeling, key for this study, has been developed. Parallel computing has been performed at the Computation Center of Novosibirsk State University, the SB RAS Siberian Supercomputer Center, the Joint Supercomputer Center of RAS, and the Supercomputing Center of the Lomonosov Moscow State University. I.S. and M.Kh. also acknowledge the RFBR project 20-02-00520. M.Kh. acknowledges support from the projects I2939-N27 and S11606-N16 of the Austrian Science Fund (FWF). M.Kh. acknowledges grant number 075-15-2019-1875 from the government of the Russian Federation under the project called ``Study of stars with exoplanets''. We thank the anonymous referee for the useful comments that led to improve the manuscript. 
\end{acknowledgements}

%
%

\end{document}